\documentclass{iau}

\usepackage{amsmath}
\usepackage{graphicx}
\usepackage{multirow}

\newcommand{\Ms}{\ensuremath{\rm M_{\odot}}}

\newcommand{\bpop}{\textsc{B-pop~}}

\newcommand{\dragonii}{\textsc{Dragon-II~}}

\newcommand{\yrgpc}{~{\rm yr}^{-1}~{\rm Gpc}^{-3}}

\begin{document}

\lefttitle{Cristiano Ugolini}
\righttitle{Piling up in the darkness: Features of the BBH mass distribution from isolated binaries}

\jnlPage{xxx}{xxx}
\jnlDoiYr{xxx}
\doival{xxxxx}

\aopheadtitle{IAU Symposium 398/MODEST-25}
\editors{xxx}
\title{The assembly of intermediate black holes with complementary approaches: Dragon II and BPop}

\author{Cristiano Ugolini}
\affiliation{Gran Sasso Science Institute (GSSI),                L'Aquila (Italy), Viale Francesco                  Crispi 7\\
            INFN, Laboratori Nazionali del Gran Sasso, 67100 Assergi, Italy\\}

\begin{abstract}
Intermediate-mass black holes (IMBHs) occupy the $10^2$--$10^5\,M_\odot$ range, but their existence remains poorly constrained. Only a few candidates have been suggested in dwarf galaxies, globular clusters, and LIGO-Virgo-Kagra detections. To investigate their formation and demographics, we adopt two complementary approaches. We first analyze the \textsc{dragonii} direct $N$-body simulations, which follow clusters with up to $10^6$ stars, capture IMBHs growth. We then employ the semi-analytic code \textsc{bpop}, calibrated on \textsc{dragonii}, to explore a broad range of cluster and cosmological conditions. 

Our models reproduce merger rates consistent with GWTC-3, with $\sim30$--60\% of BBHs forming dynamically, mainly in globular and nuclear clusters. About 2--3\% of mergers involve an IMBH, producing intermediate-mass ratio inspirals. The IMBH mass distribution spans $2.5\times10^2$--$4\times10^4\,M_\odot$, with rare growth beyond $10^6\,M_\odot$. Formation efficiency rises with initial binary fraction but declines in metal-rich environments. IMBHs thus emerge as a distinct population bridging stellar and supermassive black holes.
\end{abstract}


\maketitle

\section{Introduction}

Intermediate-mass black holes (IMBHs) occupy the $10^2$–$10^5\Ms$ range, potentially bridging stellar-mass and supermassive black holes (SMBHs) \citep[see w.g. a review by][]{ASKAR2024149}. Observational evidence is scarce: a few candidates with $\sim (5\times10^4$–$10^6)\Ms$ have been found in low-mass AGN and dwarf galaxies \citep[e.g.][]{2018ApJ...863....1C}; a dozen putative IMBHs of $10^3$–$10^4\Ms$ have been suggested in Galactic and extragalactic GCs, though most remain inconclusive \citep[see e.g.][]{2017Natur.542..203K,2022A&A...661A..68T}; and $\sim 10$ objects of $(100$–$200)\Ms$ were reported by LIGO–Virgo–Kagra (LVK) from BBH mergers, including possibly the first IMBH–IMBH binary \citep{2025arXiv250708219T}.

Several, non-mutually exclusive, formation channels have been proposed, particularly in dense clusters: (i) repeated stellar collisions \citep[e.g.][]{2002ApJ...576..899P,2023MNRAS.526..429A,2024MNRAS.531.3770R}, (ii) tidal disruption and accretion of stars onto stellar BHs \citep{2015MNRAS.454.3150G,2023MNRAS.526..429A}, and (iii) hierarchical BH mergers \citep{2002MNRAS.330..232C,2019PhRvD.100d3027R,2024A&A...690A.106M,2024MNRAS.531.3770R}. The efficiency of the formation channel depends on several uncertain cluster properties, which also shape the BBH merger population. Thus, any viable IMBH model must reproduce GW-inferred BBH merger statistics.  

A self-consistent way to investigate this topic would be through direct $N$-body simulations, but these remain computationally expensive and are limited to a few million particles. Nonetheless, the \dragonii suite \citep{2024MNRAS.528.5119A,2023MNRAS.526..429A,2024MNRAS.528.5140A} demonstrates that IMBHs can form through all three main channels.\par

Semi-analytic codes provide an alternative, capturing key dynamical processes while enabling large BBH catalogues at low cost \citep{2020MNRAS.492.2936A,2021Symm...13.1678M, 2023MNRAS.520.5259A,2024PhRvD.110d3023K, 2024A&A...688A.148T}. Here we adopt the \bpop (Binary merger POPulations, see \citealt{2019MNRAS.482.2991A, 2020ApJ...894..133A,2023MNRAS.520.5259A}, \citeauthor[][in preparation,]{Arcasedda_BPOP}) to model cosmic BBH mergers from isolated binaries (IBs) and from dynamical channels (young, globular, and nuclear clusters; YCs, GCs, and NCs).

\section{\bpop}

The \bpop code is a semi-analytic software to model BBH mergers forming from isolated stellar binaries (IBs) and dynamical interactions in YCs, GCs, and NCs. It includes prescriptions for cosmic star formation and metallicity evolution, star cluster initial masses and sizes, and their long-term evolution. The code also models the formation of stellar black holes, their natal spins and kicks, the properties of dynamically formed binaries, and the masses of black holes assembled through stellar collisions and interactions. It also handles multiple generation mergers in star clusters using numerical relativity fitting formulae to calculate the merger remnant mass and spin
\citep{2017PhRvD..95f4024J} and gravitational recoil \citep{2007PhRvL..98w1102C,2012PhRvD..85h4015L}.
We refer the interested reader to \cite{2020ApJ...894..133A, 2023MNRAS.520.5259A}, and  \citeauthor[][in preparation,]{Arcasedda_BPOP} for further details about the code workflow and features.

\section{BBH merger rate}
\label{sec: Results}

\begin{figure*}
    \centering
    \includegraphics[scale=0.2]{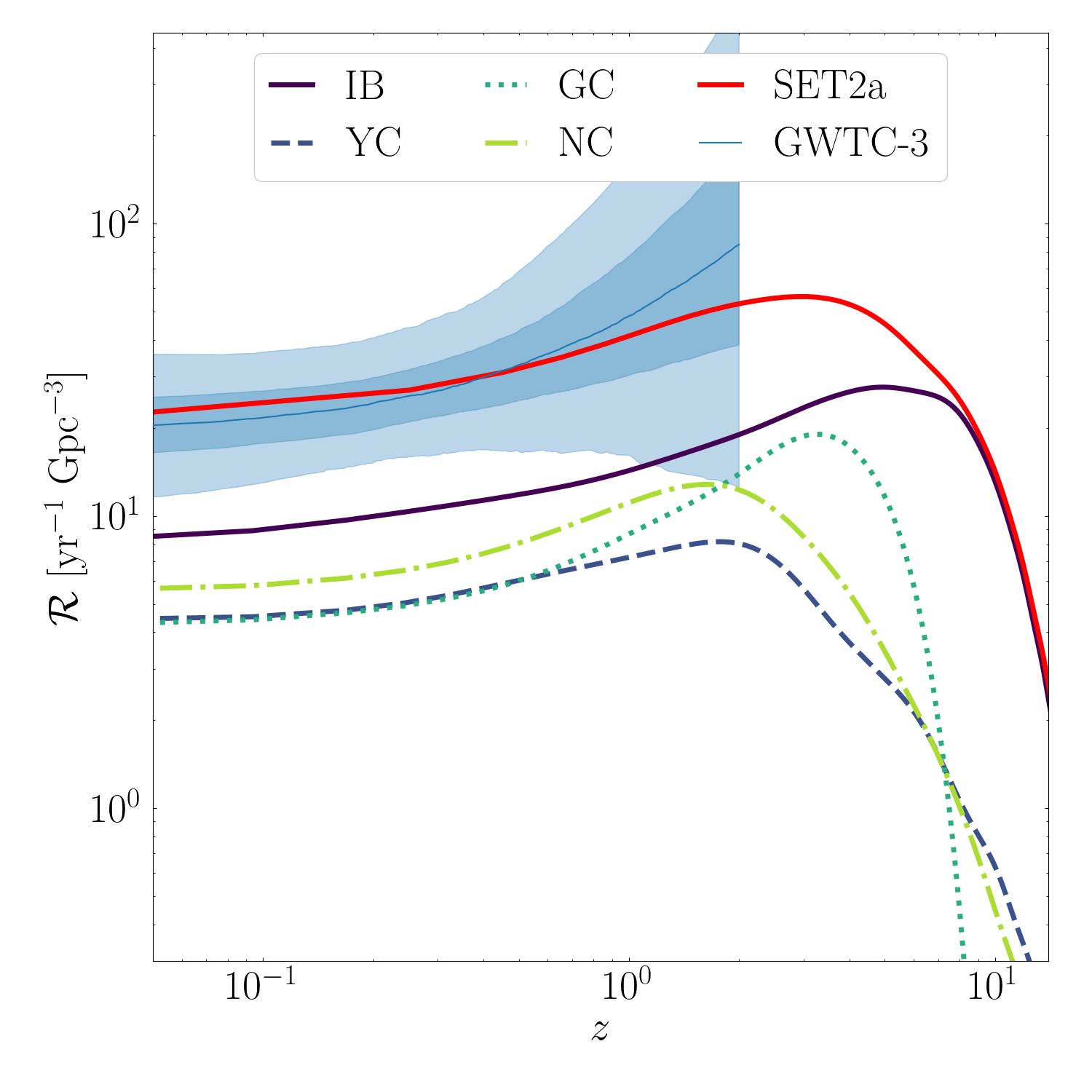}
    \caption{Merger rates as a function of redshift for the fiducial model with $f_{\rm mix} = 0.5$, $f_{\rm YC}=0.005$, and $f_{\rm IB}=1$. Different colors and line styles correspond to different formation channels: isolated binaries (IB, straight purple line), young clusters (YC, dashed blue line), globular clusters (GC, dotted green line), and nuclear clusters (NC, dot-dashed light green line). The thick red line represents the total merger rate, while the light blue line and shaded region represent the BBH merger rate inferred from the GWTC-3.}
    \label{fig:rate}
\end{figure*}

Figure \ref{fig:rate} compares the observed and simulated merger rates for our fiducial model with binary fraction at birth $f_{\rm mix}=0.5$. In what we refer to as the EB19+MF17 model, we adopt the star formation histories from \citet{2019MNRAS.482.4528E} for GCs and from \citet{2017ApJ...840...39M} for GCs/NCs. The figure shows both the total merger rate and the contribution of each channel. The former is broadly consistent with observation limits. At the fiducial redshift of $z=0.2$, we find a local merger rate $\mathcal{R}_{\rm sim} \simeq 27.1\yrgpc$, in the ballpark of values inferred from GWTC-3 observations. According to our model, around $\sim 30-60\%$ of the cosmic population of BBH mergers have a dynamical origin. For the "a" models (i.e. those adopting the EB19+MF17 star formation history; see \citeauthor[][in preparation,]{Arcasedda_BPOP} for details), most dynamical mergers form in GCs ($46-51\%$) and NCs ($30-40\%$). The relatively small contribution from BBH formed in YCs is mostly due to their short lifetime, tipically $\lesssim 0.1-1$ Gyr.

\section{IMBH sub-population}
\label{sec: Discussion}

We now focus on binaries involving an IMBH ($m>10^2\Ms$), formed either through stellar collisions, repeated BH mergers, or both. These systems, also referred to as intermediate-mass ratio inspirals (IMRIs), provide key insights into IMBH demographics.

\begin{figure*}
    \centering
    \includegraphics[width=0.7\textwidth]{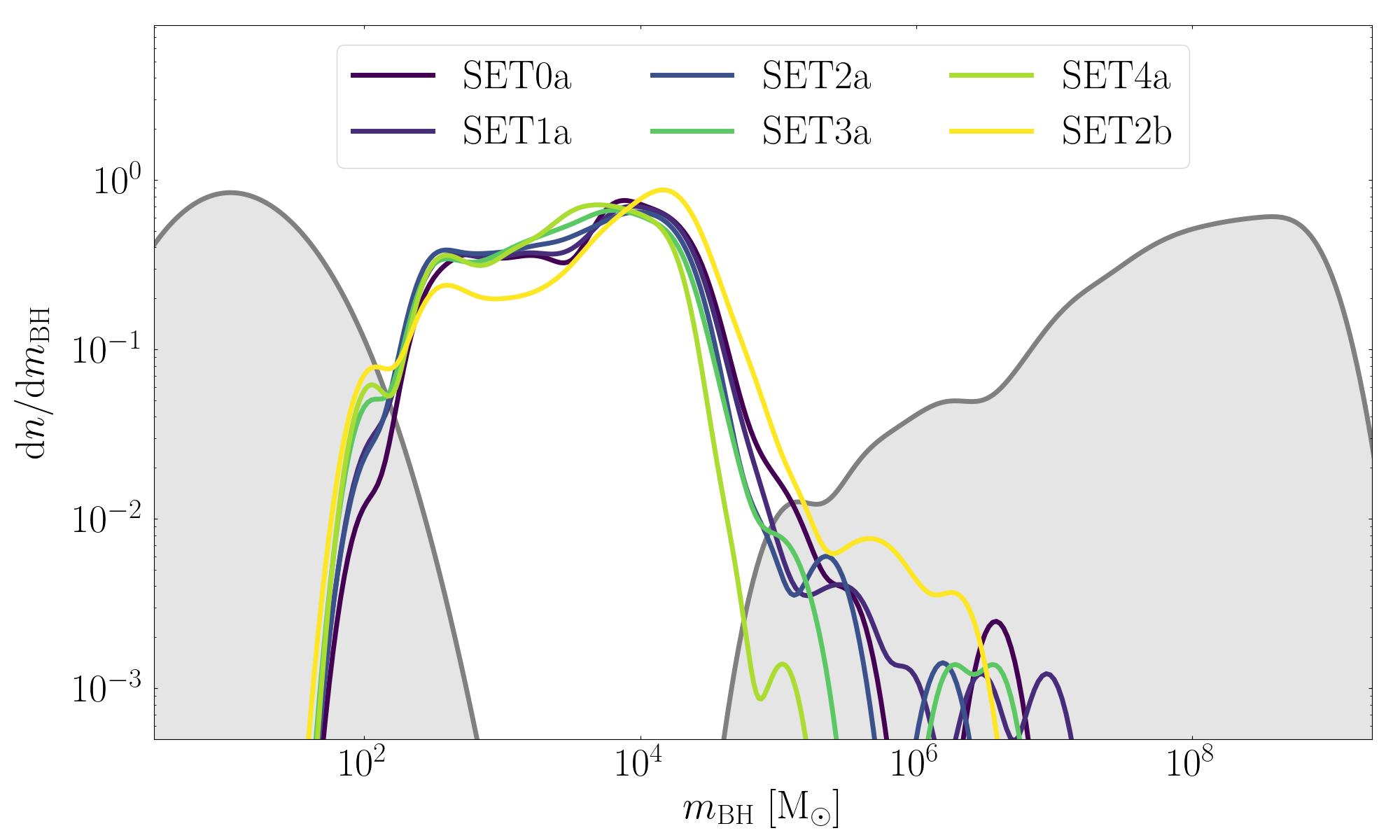}    
    \caption{IMBH mass distribution for all models. The number in the model name indicates the binary fraction at birth, with 0, 1, 2, 3, and 4 corresponding to $f_{\rm mix} = 0, 0.25, 0.5, 0.75,$ and $1$, respectively. The letter (a or b) indicates the different GCs/NCs star formation histories (see \citeauthor[][in preparation]{Arcasedda_BPOP} for details). The stellar BH distribution is obtained from our catalogs, while the SMBH distribution is based on low-$z$ observations \citep{2009ApJ...690...20S}. Areas are normalized to unity for comparison.}
    \label{fig:IBHori1}
\end{figure*}

The simulated IMBH population spans $m_{\rm IMBH} \sim (2.5\times10^2$–$4\times10^4)\Ms$, with $\sim5\%$ of objects growing to $10^6$–$10^7\Ms$, thus bridging stellar BHs and SMBHs (Figure~\ref{fig:IBHori1}). A fraction $f_{\rm IMBH}\simeq 0.018$–$0.028$ of all BBH mergers involve an IMBH, with the occurrence increasing for larger $f_{\rm mix}$ but only weakly sensitive to the assumed cluster star formation history (see \citeauthor[][in preparation,]{Arcasedda_BPOP} for further details). The high-mass tail is also enhanced at larger $f_{\rm mix}$. The environment strongly affects IMBH growth. Seeds of $\gtrsim500\Ms$ formed via stellar collisions in clusters with $M>10^5\Ms$ and $\rho>3\times10^5\Ms$ pc$^{-3}$ typically double their mass. More massive NCs ($M>10^7\Ms$, $\rho>10^6\Ms$ pc$^{-3}$) favour IMBH growth via successive BH mergers. Overall, most IMBHs originate from massive stellar collision products ($\gtrsim 97\%$), while the fraction seeded by stellar BHs rises modestly with $f_{\rm mix}$ (from $<1\%$ at $f_{\rm mix}=0$ to $\sim3\%$ for SET2b). Furthermore, metallicity plays a key role. In low-metallicity clusters ($Z<0.001$), IMRIs constitute $\sim10\%$ of mergers, whereas in metal-rich environments—where PISNe suppress massive stellar seeds—this fraction drops to $\lesssim1\%$. 

Finally, increasing $f_{\rm mix}$ reduces the maximum IMBH mass (by $\sim30\%$ from $f_{\rm mix}=0$ to 1) while raising the overall IMBH merger fraction, especially in NCs. This reflects enhanced retention of low-mass BHs with small natal kicks, which preferentially form unequal-mass binaries producing remnants more easily retained in clusters.

\section{Summary and conclusions} 
\label{sec: summary and conclusions}

We combined direct $N$-body simulations from \dragonii with semi-analytic studies performed with \bpop to investigate the assembly of IMBHs in dense stellar systems. This hybrid framework exploits the precision of $N$-body models and the statistical power of semi-analytic approaches. 

Our fiducial models reproduce local BBH merger rates consistent with GWTC-3, with $\sim30$–60\% of mergers of dynamical origin. A fraction $f_{\rm IMBH}\simeq0.02$–0.03 of all mergers involve an IMBH, producing IMRIs. The simulated IMBH mass spectrum continuously bridges stellar and supermassive BHs, peaking at $2.5\times10^2$–$4\times10^4\Ms$ with a small high-mass tail reaching $>10^6\Ms$. We find that IMBH demographics depend strongly on cluster environment and metallicity. Larger binary fraction values enhance the IMBH merger fraction but lower the maximum IMBH mass, while low-metallicity clusters sustain IMRI fractions of $\sim10\%$, compared to $\lesssim1\%$ in metal-rich systems.  
Overall, our results support IMBHs as a distinct population shaped by cluster dynamics, binary fraction, and metallicity. The combined use of \dragonii and \bpop provides a flexible framework to interpret GW observations of IMRIs and to constrain IMBH demographics across cosmic history.  

\begin{acknowledgements}
The author acknowledges Manuel Arca Sedda for useful discussion and comments. 
\end{acknowledgements}

\bibliographystyle{mnras}
\bibliography{main}

\end{document}